\documentclass[preprint2]{aastex}

\newcommand{\gr}{$^\circ$}
\newcommand{\kms}{km~s$^{-1}$}

\newcommand{\sm}{M$_\odot$}

\newcommand{\nii}{[\ion{N}{2}]}
\newcommand{\oii}{[\ion{O}{2}]}
\newcommand{\oiii}{[\ion{O}{3}]}

\newcommand{\CS}{{\small\em S}}
\newcommand{\CN}{{\small\em N}}
\newcommand{\CW}{{\small\em W}}
\newcommand{\CE}{{\small\em E}}

\slugcomment{Submitted to ApJ}

\shorttitle{The outflow of CH~Cyg} 
\shortauthors{Corradi et al.}

\begin{document}


\title{The large-scale ionized outflow of CH~Cygni\altaffilmark{1}}

\altaffiltext{1}{Based on observations obtained at the 4.2m~WHT and
2.6m~NOT telescopes operated on the island of La Palma by the ING and
the NOTSA, respectively, in the Spanish Observatorio del Roque de Los
Muchachos (ORM) of the Instituto de Astrof\'\i sica de Canarias, and
with the NASA/ESA Hubble Space Telescope, obtained at the Space
Telescope Science Institute, which is operated by the Association of
Universities for Research in Astronomy, Inc. under NASA contract No.\
NAS5-26555.}


\author{Romano L.M. Corradi}
\affil{Isaac Newton Group of Telescopes, Apartado de Correos 321,
	     E--38700 Santa Cruz de la Palma, Spain}
\email{rcorradi@ing.iac.es}

\author{Ulisse Munari}
\affil{Osservatorio Astronomico di Padova, Sede di Asiago, I-36012
Asiago (VI), Italy}
\email{ulisse@ulisse.pd.astro.it}

\author{Mario Livio}
\affil{Space Telescope Science Institute, 3700 San Martin Drive,
Baltimore, MD 21218}
\email{mlivio@stsci.edu}

\author{Antonio Mampaso \& Denise R.  Gon\c calves}
\affil{Instituto de Astrof\'{\i}sica de Canarias, c. Via Lactea S/N, 38200
       La Laguna, Tenerife, Spain.}
\email{amr@ll.iac.es, denise@ll.iac.es}

\and

\author{Hugo E. Schwarz}
\affil{Cerro Tololo Inter-American Observatory,
AURA, Casilla 603, La Serena, Chile}
\email{hschwarz@ctio.noao.edu}




\begin{abstract}

HST and ground-based \oii\ and \nii\ images obtained in the period 1996--1999
reveal a complex, ionized optical nebula around the symbiotic
binary CH~Cyg extending out to 18$''$ or 5000~A.U. from the central stars.

The observed velocity range of the nebula, derived from long-slit echelle
spectra, is 130~\kms. In spite of its complex appearance, the velocity
data show that the basic morphology of the inner regions of the optical
nebula is that of a bipolar outflow extending nearly along the plane of the
sky out to some 2000~A.U. from the center.

Even if the extension of this bipolar outflow and its position angle are
consistent with those of the radio jet produced in 1984 (extrapolated to the
time of our optical imagery), no obvious optical counterpart is visible of
the original, dense radio bullets ejected by the system.  We speculate that
the optical bipolar outflow might be the remnant of the interaction of the
bullets with a relatively dense circumstellar medium.

\end{abstract}


\section{Introduction}

One of the most debated issues in the study of the mass loss from
evolved stars is the formation of collimated outflows and jets
(cf. Livio 2000). In this respect, symbiotic stars are excellent
laboratories to study the formation and evolution of this kind of
outflows in interacting, detached binary systems.  Most of the
spatially resolved nebulae known around these stars are in
fact strongly aspherical (Corradi et al. 1999a), and highly collimated
jets have been so far identified in several systems. They are: R~Aqr
(Burgarella and Paresce 1992; Dougherty et al. 1995), He~2-104
(Schwarz et al. 1989, Corradi \& Schwarz 1993, Corradi et al. 2001),
MWC~560 (Tomov et al. 1990; Shore et al. 1994), RS~Oph (Taylor et
al. 1989), BI~Cru (Schwarz \& Corradi 1992), Hen~3-1341 (Tomov et
al. 2000), StH$\alpha$~190 (Munari et al. 2001), and CH~Cyg. In this
last system, one of the most notable jets from a symbiotic star was
detected in 1984 at $\lambda$$=$$2$~cm by Taylor, Seaquist \& Mattei
(1986, hereafter TSM86), who also measured its proper motions on a
time basis of only 75 days.

Owing to its proximity to the Earth (Hipparcos measured a distance of
268~pc, Munari et al. 1997), in CH~Cyg we have the opportunity to
study in great spatial detail the structure of a stellar jet and
follow its evolution in real time.  With this aim, we have obtained
narrowband images of CH~Cyg with the HST and the Nordic Optical
Telescope 12 to 15 years after the ejection of the radio jet. These
images, which reveal the existence of a large ionized nebula around
the symbiotic core, are presented in this paper together with
ground based echelle spectra which allow us to discuss the
spatiokinematical structure of the optical outflow and its orientation
in the sky. The observations are detailed in \S2, the images are
presented in \S3, and the kinematical data in \S4. \S5 contains a
critical review of the general properties of CH~Cyg, the discussion of
the origin of its large-scale ionized nebula and its possible link
with the 1984 radio jet.

\section{Observations}

\nii\ and \oii\ images of CH~Cyg were obtained under sub-arcsec seeing
conditions at the 2.6m~Nordic Optical Telescope (NOT) of the ORM, La Palma,
using the BroCam2 direct camera in 1996 and the ALFOSC instrument in
1997. Both cameras used a Loral 2k x 2k CCD with a scale of 0$''$.11 and
0$''$.19 per detector pixel, respectively. With ALFOSC, a coronographic plate
was used for the longest exposure, to stop light from the bright central
source and prevent saturation of the CCD.  The central wavelength and FWHM of
the filters used at the NOT are: \oii\ (372.5/2.9~nm) and \nii\
(658.9/0.9~nm).  In the following, we will therefore indicate with [OII] and
[NII] the emission in the nebular lines \oii$\lambda$$\lambda$372.7,372.9~nm,
and \nii$\lambda$658.3~nm.  Given the very similar ionization potentials of
these oxygen and nitrogen ions, we will also assume that they are emitted in
similar regions of the ionized nebula around CH~Cyg. Exposure times
and seeing are detailed in Table~\ref{T-logobs}.

\oii\ images of CH~Cyg were also obtained with the Hubble Space Telescope in
1999, using the STIS camera and aperture F28$\times$50\,OII. The \oii\ filter
of STIS is centered at 374.0~nm and has a FWHM of 8.0~nm.  We obtained a short
image of CH~Cyg (200~sec split into two repeated exposures to allow for
cosmics removal and limit saturation of the central source) as well as a
deeper one (2508~sec split into 3 exposures).  The STIS CCD pixel size is 
0$''$.05, and stars have a FWHM of about 1.4 pixels at the \oii\ wavelengths.

The kinematics of the ionized nebula around CH~Cyg was studied by
means of long-slit, echelle \nii\ spectra obtained in 1997 at the
2.6m NOT and at the 4.2m William Herschel Telescope (WHT) of the ORM.
At the NOT, we used the IACUB spectrograph, provided with a Thompson
THX31156 CCD which gives a spatial scale of 0$''$.14~pix$^{-1}$. To
increase the signal--to--noise ratio, 2$\times$2 binning was done
along both the spatial and spectral directions.  The projected slit
width was 0$''$.65, providing a spectral resolution $R=\lambda /
\Delta\lambda=$30000, with a reciprocal dispersion of 0.009~nm per
binned pixel.  At the WHT, we used the UES spectrograph equipped with
a SIT1 CCD, providing a spatial scale of 0$''$.36~pix$^{-1}$.  The
spectral resolution was $R=50000$, with a reciprocal dispersion of
0.007~nm per pixel and a slit width of 1$''$.0.  The spectrograph slit
was positioned both on the central star, and offset to cover most of
the nebular features of CH Cyg. Exposure times, slit position angles
and offsets are summarized in Table~\ref{T-logobs}.

\section{Morphology}

The \nii\ and \oii\ images of CH~Cyg are presented in
Figure~\ref{F-ima}.  These images reveal for the first time the
existence of a large ionized nebula extending out to 18$''$ (or
$\sim$5000 A.U. for a distance of 268~pc) from the central stars. In
its outer parts (Figure~\ref{F-ima}, top panels), the nebula is
fragmentary, showing faint ``plumes'' in the NW quadrant and ``knots''
toward the East.  The bright, inner body of the nebula is also
complex, but generally elongated along the NW and SE directions
(Figure~\ref{F-ima}, middle panels). The detailed morphology of its
brightest parts is revealed by the HST images (Figure~\ref{F-ima},
bottom).  Toward the NW, a collimated `spray' of emission extends from
the central stars along P.A.=$-35$\gr\ up to a distance of 3$''$.2 and
possibly out to 4$''$.3 (1000 A.U.), with it's width increasing with
distance from $\sim$150 to $\sim$300 A.U.  On the SE side, the \oii\
emission in the innermost arcsecond from the central stars (inset box
in Figure~\ref{F-ima}, bottom right) is elongated roughly toward the
South : its properties and formation are discussed by Eyres et
al. (2001). Outwards, emission bends toward the SE opening and forming
arcs and knots which become fainter with distance, but which can be
followed out to the peripheric zones of the nebula in the deep NOT
images.

We have considered the possibility that the collimated `spray' of
emission extending NW from the central stars is an instrumental
artefact, especially because it appears along the serial read axis of
the CCD. To check it, we have constructed an empirical
point-spread-function (PSF) using archive STIS images of a calibration
star which was observed through the same \oii\ filter, in the same
position in the CCD chip, and only 17 days later than CH~Cyg. This PSF
was then subtracted from our images, after scaling its peak emission
to match that of CH Cyg (note that a small scaling factor, $\sim$2,
was applied in the case of our short image, which is only slightly
saturated). The subtracted images are presented in Figure~\ref{F-ima},
bottom right, and in the inset box.  A similar result is obtained
using the PSF to restore the images with the Lucy-Richardson
algorithm.  In both cases, while most instrumental artefacts are
significantly reduced, the NW collimated emission is instead enhanced,
making us confident that it is a real nebular structure.  

In the short subtracted image (Figure~\ref{F-ima}, inset box), a small ring
is visible which extends out to 0$''$.35 (100~A.U.)  West of the center. It
is not clear whether this ringlet is a real feature or the typical ghost
appearing in STIS images close to bright stars (see {\it
www.stsci.edu/cgi-bin/stis?stisid=224\&cat=performance\&subcat=foibles}).
Together with another dubious feature, it is marked with the label ``real?''
in Figure~\ref{F-ima}.

HST images of CH~Cyg in other filters were obtained with WFPC2 by Eyres et
al. (2001). The comparison of our \oii\ image with their \oiii\ one is
especially interesting: the latter one does not show the `spray' of
collimated outflow on the NW side of the stars, nor it shows the extended
`arcs' toward the SE except for the brightest emission in the innermost
1$''$.5. This confirms our previous finding (e.g. Corradi et al. 1999b) that
radiation from low-ionization species such as \nii\ and \oii\ is likely the
best tracer of large-scale outflows (i.e. several orders of magnitude larger
than the binary separation) in symbiotic systems.

\section{Kinematics}

The slit positions of the echelle spectra covered most of the nebular
features of CH~Cyg. In particular, the NOT spectra provide a detailed mapping
of the bright inner regions, while the WHT ones were meant to cover the faint
outer structures.  In all spectra, radial velocities were measured by
multi-Gaussian fitting of the \nii\ line profile at selected positions along
the slits, and correcting for the heliocentric systemic velocity of CH~Cyg of
$-58$~\kms\ (Skopal et al. 1989).

The number of velocity components observed in the spectra at all slit
positions reflects the morphological complexity of the nebula.  Radial
velocities span an overall range of 130~\kms. This is significantly
larger than the velocity of a standard red giant wind, implying that
fast winds or ejecta from the hot component must play a role in
producing the observed nebula, and possibly also ionizes it by
shock heating. The observed radial velocities, however, are also
clearly lower than the total expansion of 1400/$\sin i$~\kms\ derived
for the 1984 radio jet by TSM86 using its proper motions ($i$ is the
inclination of the jet to the line of sight). Either the optical
nebula is not associated with the radio ejecta, or the latter have been
dramatically slowed down between 1984 and 1997, or projection
effects are large, i.e. material is expanding nearly along the plane
of the sky.

The velocity field of the inner nebula is shown in Figure~\ref{F-notspe}.
Most \nii\ position-velocity plots are reminiscent of `ellipses', albeit
fragmentary and distorted. The spatial position and extent of the ellipses
are indicated in the bottom panels of Figure~\ref{F-notspe}, superimposed on
the HST and NOT images.  These kinematical figures are better seen in the SE
side, where the spatial size of the ellipses also appears to increase
slightly with distance from the center, suggesting that the overall outflow
has a 3-D conical or bipolar geometry.  It is clear that the aperture of the
bipolar outflow described by the kinematical ellipses ($\sim10^{16}$~cm) is
much larger than the innermost, collimated NW `spray', and also larger than
the brightest SE `arcs' seen in the HST \oii\ image.  The size of the
ellipses matches instead, within the uncertainty in the spatial `zero' point
along each slit position, that of the elongated nebulosity seen in the deep
NOT \oii\ image (Figure~\ref{F-ima}, top right, and ~\ref{F-notspe}, bottom
right). The sharp SE `arcs' and the NW `spray' of collimated emission seen in
the HST image might then be surface brightness enhancements on the walls of a
wider bipolar outflow. As with velocities, the analysis is limited by the
irregular shape of the kinematical figures, but most ellipses have a
kinematical axis of between 100 and 120~\kms, and all them are fairly well
centered on the adopted systemic velocity. This latter property indicates
that the axis of the bipolar outflow lies almost exactly in the plane of the
sky.  Under this hypothesis, and also assuming that velocities are directed
radially from the central stars, deprojected expansion velocities result to
be around 100~\kms, i.e. not dramatically larger than the observed
ones. These are still notably lower than those of the 1984 radio jet. The
kinematical age of this bipolar outflow would result to be 25~yrs for the
material 1$''$.5 SE of the star (400~A.U.), increasing to 75~yrs for gas at a
distance of 5$''$ (1300~A.U.). The figures above, however, should be taken
with caution because of the several simplistic assumptions involved. If, for
instance, velocities were not directed radially and gas were instead flowing
along the walls of the bipolar outflow, then deprojected velocities would be
much larger, and be eventually comparable with those derived from the
apparent expansion of the radio jet. Kinematical ages would also be shorter,
especially for gas at $\ge$3$''$ SE of the center, where the flow looks
almost cylindrical with the walls parallel to the plane of the sky.

The radial velocities of the faint outer nebula from the WHT spectra
are indicated in Figure~\ref{F-outervel}. Radial velocities in the
plumes and knots span a range from $-72$~\kms\ to $+55$~\kms\ with
respect to the systemic velocity. Again, the kinematics indicates that
the geometry of the outer nebula is complex and asymmetrical, since
most components are redshifted regardless of which side of the nebula
they are located on. Note that this could imply reflection off these
features instead of emission from them (cf. M2--9 in Schwarz et al.1997).

\section{Discussion}

\subsection{Basic parameters of the system}

Before discussing the origin of the extended nebula of CH~Cyg, some
discussion about the basic properties of its central stars is in
order.  \objectname{CH~Cyg} (\objectname{HD182917} = \objectname{HIC
95413}) is one of the best studied symbiotic stars, with $\sim$550
publications related to it.  Before 1963, it was known as a low
amplitude variable with a weak periodicity around 100 days and no
spectral peculiarities; CH~Cyg was in fact adopted as a standard star
for the spectral type M6III in the MKK system. Starting in 1963, the
presence of a hot companion was revealed by the sudden appearence of
emission lines, a marked brightening in the blue and UV, and irregular
flickering at short wavelengths with a large amplitude and timescale
of minutes (Deutsch 1964, 1967; Cester 1969; Walker et
al. 1969). CH~Cyg was consequently reclassified as a symbiotic star.

This first outburst ended in 1970, and was followed by a longer and more
pronounced one ($V$ up to 5.5~mag) lasting from 1977 to 1986, which was
widely studied at all wavelengths. Toward its end, in the summer of 1984, the
drop in brightness and the disappearance of the enhanced blue/ultraviolet
continuum was accompanied by the ejection of a high velocity radio jet and a
sudden broadening of the emission lines (Tomov 1984, Selvelli \& Hack 1985,
TSM86).  In the following years, the brightness of CH~Cyg remained close to
quiescent levels with moderate re-brightenings in 1993-1994 and 1998-2000
(Taranova \& Shenavrin 2000).

As with the cool component of CH~Cyg, the study of the UBV lightcurve from
1885 to 1988 (Mikolajewski et al. 1990) and of JHKLM photometric data for the
last 25~yrs (Munari et al. 1996, Taranova \& Shenavrin 2000) have revealed
the following properties: {\it i)} low-amplitude optical and near-IR
variability with several periodicities (100, 770 1300, and 1980 days); {\it
ii)} a secular decrease in the mean luminosity of the red giant
($\bigtriangleup V \geq 3-4$ mag over 120~yrs, with a rapid acceleration in
the last 40~yrs); {\it iii)} variable obscuration from several episodes of
dust condensation in the giant wind; and {\it iv)} infrared properties
typical of the bulge/thick-disk population of the Galaxy, favouring a red
giant mass around 1~\sm.

The hot component of CH~Cyg is quite unusual for symbiotic stars, having
never reached temperatures $\geq$100,000~K that power the typical high
ionization emission line spectrum of symbiotic stars.
During activity, the IUE ultraviolet spectrum of CH~Cyg resembles that of an
A star (Hack \& Selvelli 1982), although no model atmosphere gives an
acceptable fit to the erratically variable UV continuum.  At quiescence, the
UV continuum resembles instead that of a moderately hot ($\sim$35,000~K)
white dwarf (e.g. Mikolajewska et al. 1993). In spite of the relatively low
temperature of its hot component, CH~Cyg shows strong X-ray emission which is
variable at long and short timescales, and has two main broad components
peaking at 0.8~keV and at 4-5~keV both originating in optically thin hot
plasma (Ezuka et al. 1999, see also Leahy \& Taylor 1987 and Murset et
al. 1997).

A robust distance determination for CH~Cyg was provided by Hipparcos:
$d = 268$~pc $\pm23\%$ (Munari et al. 1997).  Its large barycentric
radial velocity supports the association of CH~Cyg with the spheroidal
or thick-disk component of the Galaxy as suggested by the infrared
properties.

The most accepted orbital period for the CH~Cyg symbiotic system is
around 15.5~yrs with hints of an eclipsing nature (Mikolajewski et
al. 1990), although some debate has been raised in recent years
following the claim by Hinkle et al. (1993) that the system might be
triple, with 2 and 15.5~yrs orbital periods, a claim not supported by
the analysis of Munari et al. (1996).

\subsection{The origin of the optical nebula}

Prior to our observations, the only evidence for spatially resolved outflows
from CH~Cyg was the spectacular radio jet discovered by TSM86, subsequent
radio imagery by Crocker et al. (2001), and the optical spectroscopy of Solf
(1987) and Tomov et al. (1996) limited to the innermost arcsecond.  With our
optical data, in principle we have the opportunity of exploring how the jet
has evolved 12 to 15 years after its production.  According to the expansion
rate measured by TSM86 (0$''$.55~yr$^{-1}$ on each side of the central star),
at the time of our optical imagery the jet should have reached a distance
from the central source of 7$''$ (1996 and 1997, NOT imaging) and 8$''$
(1999, HST images), i.e. large enough to be easily resolved even in our
ground-based images (see the crosses in Figure~\ref{F-ima}, top/middle-right
and bottom-left, and in Figure~\ref{F-outervel}).

Following TSM86, the jet was essentially made up of two components, each with
a size of $\sim$10$^{14}$~cm and a density $\ge2\times 10^6$~cm$^{-3}$.
Because of the interaction with the ambient medium and the radiation field,
such supersonic, dense `bullets' will expand, and eventually will be
fragmented and disrupted by Kelvin--Helmholtz and, more importantly, by
Rayleigh--Taylor instabilities.  The disruption timescale can be estimated
using the formulae in Jones, Kang, \& Tregillis (1994) and Klein, McKee, \&
Colela (1994).  For environmental densities of $\sim$10$^2$~cm$^{-3}$, i.e. a
density contrast with the jet of 10$^4$, the time at which half of the bullet
material will be mixed up with the ambient medium is on the order of
80~yr. For larger environmental densities, namely $\sim$10$^3$~cm$^{-3}$ and
$\sim$10$^4$~cm$^{-3}$, the timescales are 30~yrs and 8~yrs, respectively.
In the same time interval, the bullets are expected to be slowed down by ram
pressure only by some 25$\%$ of their initial speed.  Considering the figures
above, and all the uncertainties in the physical parameters of the jet and
the ambient medium (especially the density of the latter which is presently
not known), we cannot therefore rule out completely the possibility that at
the time of our optical imaging the radio jet has been disrupted and has
mixed up with the circumbinary gas. Environmental gas densities of more than
10$^3$~cm$^{-3}$ are in fact possible in the inner circumbinary regions.
This might be the reason why no obvious optical counterpart of the original
bullets of the radio jet is seen in our optical images. Note that VLA
radio maps at $\lambda$$=$$2$~cm taken on March 20, 1986, i.e. 1.2~yr after
the images of TSM86, show the bullets at approximately the correct position
as predicted according to their expansion rate, while they are not visible
any longer in maps taken another 2.6~yrs later (Crocker et al. 2001).

There are, however, nebular structures which might be linked with the
evolution of the radio jet.  The symmetry axis of the inner bipolar outflow
of CH~Cyg discussed in the previous section is in fact oriented as the radio
jet along P.A.$\sim-45$\gr. Also the overall size of this bipolar outflow
(Figure~\ref{F-ima}, top-right) is roughly the one expected by extrapolating
the proper motions of the 1984 radio jet.  We tentatively propose that the
optical bipolar nebula might be the result of the interaction of the original
dense bullets with the circumstellar gas and radiation field from the hot
component. Although the details of the process might be complex, it is clear
that the high velocity bullets will sweep up the circumstellar material in
the direction of their motion forming large bow-shocks, and at the same time
will expand, dilute, and be disrupted by dynamical instabilities (cf. also
Mellema et al. 1998; Soker \& Regev 1998).  Some of the features seen in the
\oii\ HST image (such as the inner `arcs' and `spray') might in fact be
tentatively associated with such instabilities.  Detailed models are needed
to test whether these processes together can produce, with the correct space
and time scales, the bipolar `cavity' observed in the optical forbidden lines
in CH~Cyg.

Whatever is the relation between the radio jet and the optical nebula,
our spatiokinematical analysis indicates that the bipolar outflow of
CH~Cyg lies almost exactly in the plane of the sky. As CH~Cyg is an
eclipsing binary, i.e. the orbit is seen edge-on, our results are
fully consistent with the idea of a {\em polar} outflow.  This is an
important finding, since albeit predicted by theories (e.g. Morris
1987, Soker 1997), direct evidence that collimated outflows from wide
interacting binaries occur along the polar direction of the orbits
can hardly be found.

\section{Summary and conclusions}

The HST and NOT narrowband images and spectra of CH~Cyg have revealed
a complex, ionized nebula around this enigmatic symbiotic system,
which extends 5000~A.U. from the central system and has an observed
velocity range of 130~\kms. The spatiokinematical analysis suggests
that the kinematical age of the whole nebula (including the outermost
plumes and knots) is certainly $<100$~yrs, and possibly much lower.

In the optical images, no obvious counterpart of the radio bullets
ejected 12-15~yrs before is visible, but we tentatively interpret the
observed bipolar optical outflow as being the remnant of the
interaction of the original jet with a relatively dense circumstellar
medium, based on the fact that their projected alignment is the same
and the size of the optical outflow is consistent with the proper
motions originally measured for the radio bullets.  Further imagery
over the next years will be important to test this hypothesis and
better understand the evolution of such a short-lived, nearby stellar
jet.

The data are also consistent with the idea of a collimated flow along
the polar axis of the orbit.  The determination of the position angle
in the sky of the binary axis of CH~Cyg, e.g. by means of
spectropolarimetric observations (cf. Schmid et al. 2000), would
finally confirm it.

\acknowledgements

ML acknowledges support from NASA Grant GO~7378 from the Space
Telescope Science Institute, which is operated by AURA, Inc., under
NASA contract NAS 5-26555.  The work of AM, DRG and RLMC is supported
by the Spanish grant DGES PB97-1435-C02-01.



\clearpage



\clearpage

\begin{figure*}
\caption{The NOT and HST images of CH~Cyg, on a logarithmic intensity
scale. North is at the top, East to the left: in order to have this
orientation, the original HST images have been rotated counterclockwise 
by 54\gr\
. Details of the nebular core are given in the inset
box at the bottom right, which is enlarged $\times$2 compared to the
other HST images. In the NOT and HST \oii\ images (top/middle-right
and bottom-left), we have marked with crosses the positions at which
the 1984 radio jet would be expected at the epoch of our optical imagery,
according to the proper motions measured by TSM86.}
\label{F-ima}
\end{figure*}

\begin{figure*}
\caption{At center, the location of the slits for the NOT
spectroscopy. Above and below it, the \nii\ spectra, on a logarithmic
scale unless otherwise indicated. The spatial direction is the
vertical one (the size of each box is 15 arcsec, and corresponds to
the size of the slits drawn over the image), while wavelengths
increase along the horizontal direction.  At bottom, the ellipses
drawn onto the \oii\ HST and NOT images represent the location and
extent of the kinematical ellipses outlined by the spectra (see
text).}
\label{F-notspe}
\end{figure*}

\begin{figure*}
\caption{The observed radial velocities, as derived by Gaussian fitting in
the WHT spectra and corrected for the systemic velocity, for the faint outer
features of the nebula of CH~Cyg.}
\label{F-outervel}
\end{figure*}



\begin{table*}
\caption[]{Log of the observations}
\begin{tabular}{lrlll}
\multicolumn{5}{l}{\it\bf Images}\\
\hline
Instrument & date     & filter    & exp. time (sec) & seeing \\
\hline\\[-2pt]
NOT+Brocam2 &  4.6.1996 & \nii & 20, 90, 300, 1200      & 0$''$.6-0$''$.8\\
NOT+ALFOSC  & 14.7.1997 & \oii & 60, 240$\times$5, 1800C$^\star$ & 1$''$.0\\
HST+STIS    & 1.10.1999 & \oii & 200, 2508      & \\[5pt]
\end{tabular}

$^\star$ C = Coronographic plate\\[-3
pt]

\begin{tabular}{llrlrc}
\multicolumn{6}{l}{\it\bf Long--slit spectra}\\
\hline
Instrument & date   &P.A.(\gr)& offset$^\ast$ & exp. time & seeing\\
\hline\\[-2pt]
WHT+UES  & 10.7.1997 & $+$45 & centered          &  600 & 2$''$.0  \\
	 &	     & $-$40 &  1.1\CW,  1.0\CS & 1800 &  \\
         &           & $-$50 &  3.9\CW,  4.6\CS & 2400 &  \\
         &           & $+$30 &  8.5\CW,  7.0\CN & 2000 &  \\
         &           & $+$45 &  2.1\CE,  2.1\CS &  900 &  \\
         &           & $+$45 &  3.0\CE,  3.0\CS & 1200 &  \\
         &           & $+$5  & 10.5\CW,  0.9\CN & 1800 &  \\
         &           & $+$60 &  1.9\CW,  3.2\CN & 1200 &  \\
         &           & $+$95 &  1.0\CE, 11.5\CS & 2400 &  \\
NOT+IACUB& 17.7.1997 & $+$45 &  2.8\CN,  2.8\CW\, ({\bf a})    	&600&0$''$.7\\
         &           & $+$45 &  2.1\CN,  2.1\CW\, ({\bf b})    	& 1800 & \\
         &           & $+$45 &  1.4\CN,  1.4\CW\, ({\bf c})    	&  900 & \\
         &           & $+$45 &  0.7\CN,  0.7\CW\, ({\bf d})    	& 1800 &  \\
         &           & $+$45 &  0.7\CS,  0.7\CE\,\,\, ({\bf e}) &  900&  \\
         &           & $+$45 &  1.4\CS,  1.4\CE\,\,\, ({\bf f}) &  900&  \\
         &           & $+$45 &  2.1\CS,  2.1\CE\,\,\, ({\bf g}) & 1200&  \\
         &           & $+$45 &  3.0\CS,  3.0\CE\,\,\, ({\bf h}) & 1800&  \\
	 & 18.7.1997 & $+$45 &  3.8\CS,  3.8\CE\,\,\, ({\bf i})&1800&0$''$.7\\
\end{tabular}

$^\ast$Slit offsets are given in arcseconds from the central star
toward North (\CN), South (\CS), East (\CE), or West (\CW). The
characters in boldface are labels used in
Figure~\ref{F-notspe}. Seeing values are FWHM and are measured
directly from the images or spectra.\\
\label{T-logobs}
\end{table*}





\end{document}